\documentclass{article}
\newif\ifmarglabs
\marglabsfalse
\def\drlabel#1{\ifmarglabs
\marginpar{\scriptsize label: \\ #1} \else \relax\fi}
\def\drcaption#1{
    \footnotesize Figure \thefigure\quad #1}
\usepackage{epsfig}

\newcommand{\sE}{{\cal E}}

\newcommand{\xze}{x_{0}}

\newcommand{\xf}{x_{{\rm f}}}

\newcommand{\Eb}{\overline{E}}
\newcommand{\xt}{x_{{\rm t}}}
\newcommand{\cd}{\partial}

\newcommand{\pad}[2]{\frac{\cd #1}{\cd #2}}

\title{A Scaling Law for the Energy Levels of a \\
Nonlinear Schr\"{o}dinger Equation}
\author{R Hasson and D Richards \\ Mathematics Faculty \\
The Open University \\
Milton Keynes MK7 6AA \\
England }
\begin{document}
\maketitle
\begin{abstract}
\noindent
It is shown that the energy levels of the one-dimensional nonlinear
Schr\"{o}dinger, or Gross-Pitaevskii, equation with the homogeneous
trap potential $x^{2p}$, $p\geq 1$, obey an approximate scaling law and
as a consequence the energy increases approximately linearly with the
quantum number. Moreover, for a quadratic trap, $p=1$, the rate of increase
of energy with the quantum number is independent of the nonlinearity: this
prediction is confirmed with numerical calculations. It is also shown that
the energy levels computed using a variational approximation do not
satisfy this scaling law.
\end{abstract}

\section{Introduction}
The Bose-Einstein condensate is described, approximately, by a mean-field
approximation, see for example Friedrich (1998), that gives the
Gross-Pitaevskii equation. In the one-dimensional problem considered here
this equation takes the form,
\drlabel{eq:1-1}
\begin{equation}
-\frac{\hbar^{2}}{2\mu}\frac{d^{2}y}{dx^{2}}+
\frac12\mu\omega^{2}x^{2}y+A|y|^{2}y=Ey,
\label{eq:1-1}
\end{equation}
where $x$ is the spatial coordinate, $\mu$ the atomic mass of the atoms
comprising the condensate, $\omega$ the classical frequency of a single atom
in the trap potential. The nonlinear parameter $A$
results from the use of a mean-field
approximation to describe the particle interactions and
is defined in terms of fundamental constants, $A=4\pi\hbar^{2}\alpha_{0}N/\mu$
where $\alpha_{0}$ is the scattering length and $N$ the effective density
of atoms along the condensate axis.
In most experimental circumstances the nonlinear constant $A$ is large
so perturbation methods are of little value. For the ground state, because
the wave function varies relatively slowly and because the nonlinearity
is large the Thomas-Fermi approximations, equation~\ref{eq:2-11} below,
provides a reasonable approximation to both the energy level and the
wave function. For excited states no such simple approximation seems to be
available. Yabulov {\em et al} (1997) have derived a re-normalised
perturbation theory that gives approximate energy levels and wave functions,
but we show in section~\ref{sec:num} that this method seems to provide
a poor estimate of the excited energy levels.

In this paper we show that the energy levels satisfy a simple approximate
scaling law and consequently that they are given approximately by the
simple formula,
\drlabel{eq:1-2}
\begin{equation}
E_{n}(A)=\frac12\left( \frac32 A\omega \sqrt{\mu}\right)^{2/3} +
\frac{7\pi}{32}\omega\hbar n.
\label{eq:1-2}
\end{equation}
The first term is just the Thomas-Fermi estimate of the ground state
energy, obtained by neglecting the kinetic energy term. The second
term is the dominant correction and is linear
in $n$ independent of $A$. We show also that the latter behaviour is a
consequence of the particular form of the trap potential.

\section{Theory}
The eigenvalues of equation~\ref{eq:1-1}, $E_{n}(A)$, $n=0,\,1,\,2\cdots$,
are those values of $E$ for which $y(x)$ satisfy the boundary conditions
$|y|\to 0$ as $|x|\to\infty$ {\em and} the normalisation condition
\drlabel{eq:2-2}
\begin{equation}
\int_{-\infty}^{\infty}dx|y(x)|^{2}=1.
\label{eq:2-2}
\end{equation}
For real eigenvalues we may assume $y(x)$ to be real.

Two of the four independent parameters in this equation may be removed by
rescaling $x$ and $y$ and ensuring that the normalisation conditions
is invariant,
\[
x=\alpha x', \quad  y=\frac{y'}{\sqrt{\alpha}}, \quad
\omega=\frac{\omega'}{\hbar}, \quad A=\alpha A', \quad
\alpha=\frac{\hbar}{\sqrt{\mu}},
\]
which replaces $\mu$ and $\hbar$ by unity. In the following we drop all
primes.

By re-writing equation~\ref{eq:1-1} in the form
\drlabel{eq:2-3}
\begin{equation}
\frac{d^{2}y}{dx^{2}}+\pad{V}{y}=0, \quad
V(y,x)= \Eb(x)y^{2}-\frac12 Ay^{4}, \quad
\Eb(x)=E-\frac12\omega^{2} x^{2}
\label{eq:2-3}
\end{equation}
and treating $x$ as the `time' we may interpret equation~\ref{eq:1-1} as that
of a classical particle of unit mass moving in a time-dependent potential,
$V(y,x)$. Conventional methods of classical dynamics provide
a means of estimating the eigenvalues.

The potential $V(y,x)$ is stationary at $y=0$ and this is a minimum for
times $x<\xze=\sqrt{2E}/\omega$ and for these times there are also maxima at
\[
y^{2}=y_{m}(x)^{2}=\Eb(x)/A.
\]
For larger times, when $\Eb(x)<0$, there is only a maximum
at $y=0$. Hence quasi-periodic motion is possible for small times
but for larger times almost all orbits diverge as
$|x|\to\infty$: for every $E>0$, however, there are initial conditions for
which $y(x)\to 0$ as $x\to\infty$.

To be specific consider the even solution with initial conditions
$y(0)=a>0$ and $y'(0)=0$. For small $a$ and large enough $E$ this orbit will
oscillate in the potential well until the barrier at $y=y_{m}(x)$ is low
enough for the orbit to either escape or to ride on the barrier top and
eventually to zero: most orbits escape to infinity. Examples of these types
of orbit are shown in the following figure. Here $E=15.0810$, $A=100$,
$\omega=1$ and $a=a_{1}=0.23975967$ and $a=a_{1}\pm0.0000001$; the converged
solution is not normalised.

\begin{center}
\begin{picture}(240,120)
 \put(0,0){\epsfig{file=hass01.eps}}
 \put(120,63){$x$}
 \put(28,105){$y(x)$}
\end{picture}

\refstepcounter{figure}
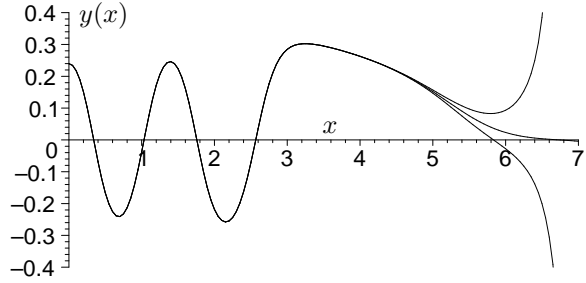
\drcaption{Some examples of even solutions of
equation~\ref{eq:1-1}, with $E=15.0810$\\and $y(0)=a$, given in the text.
\label{f:01}}
\end{center}

This figure shows that the required solutions with $y(x)\to 0$ as
$|x|\to\infty$ comprise a quasi-periodic
part, for $|x|<\xt$ where $\xt$ is defined in equation~\ref{eq:2-6} below,
and a monotonically decreasing segment for $|x|>\xt$. It also shows that
the distance between nodes is almost constant: reasons for this are discussed
later.

Consider the oscillatory region. When $\Eb=$constant it follows from the
definition of the Jacobi elliptic function that the odd and even solutions
are, respectively
\drlabel{eq:2-10}
\begin{equation}
y=a\,{\rm sn}(\tau,k), \quad (y(0)=0), \quad
y=a\,{\rm sn}(K-\tau,k), \quad (y(0)=a),
\label{eq:2-10}
\end{equation}
where $K=K(k)$ is  the complete elliptic integral of the first kind and
\[
\tau=x\sqrt{2\Eb-a^{2}A}, \quad k^{2}=\frac{Aa^{2}}{2\Eb-Aa^{2}}.
\]
The period of these oscillations is
\drlabel{eq:2-7}
\begin{equation}
T=\frac{4}{\sqrt{2\Eb-Aa^{2}}}K(k).
\label{eq:2-7}
\end{equation}
When $\Eb$ is constant the action of the above oscillatory solution may
be written in the form
\drlabel{eq:2-4}
\begin{equation}
I=\frac12 a^{2}\sqrt{2\Eb}\,F(k), \quad
F(k)=\frac{4}{3\pi k^{2}\sqrt{1+k^{2}}}
\left( (1+k^{2})E(k)-(1-k^{2})K(k)\right),
\label{eq:2-4}
\end{equation}
where $E(k)$ is the complete elliptic integral of the second kind, and is
not to be confused with the energy.
For each $\Eb$ there is bound motion if $0 < Aa^{2} < \Eb$ and as $k$
increases from zero to unity $F(k)$ decreases from 1 to
$4\sqrt{2}/(3\pi)\simeq 0.6$.
The action is bounded by $0 \leq I \leq I_{s}$, where $I_{s}$ is the action of
the bound, non-periodic motion on the separatrix, where $Aa^{2}=\Eb$ ($k=1$),
\drlabel{eq:2-5}
\begin{equation}
I_{s}=\frac{4}{3\pi A} \Eb^{3/2}.
\label{eq:2-5}
\end{equation}

Now consider the effect of $\Eb$ decreasing, but changing little during
one period of the unperturbed motion. The principle of adiabatic invariance
(Percival and Richards, 1982, chapter~9) shows that the action is almost
invariant.
The separatrix action, however, is not constant and decreases to zero at
$x=\xze$ where
$\omega \xze=\sqrt{2E}$. All orbits cease to oscillate before this time and
if the change in $\Eb$ is sufficiently slow  this change occurs when the
action equals the separatrix action. If $\xt$ is this time it is given
by the solution of
\drlabel{eq:2-6}
\begin{equation}
\frac{4}{3\pi}\left( E-\frac12 \omega^{2}\xt^{2} \right)^{3/2}=AI(E)
\label{eq:2-6}
\end{equation}
where the action is evaluated at $E$, the initial value of $\Eb$. Adiabatic
invariance shows that the solution oscillates with a local period, $T$,
given by equation~\ref{eq:2-7}, which depends upon $x$. However, the period
although singular at $\Eb(x)=Aa^{2}$, does not change significantly until
$\Eb(x)$ is close to $Aa^{2}$, so the nodes of the wave function are
almost equally spaced.

The quantum number, $n$, that labels the state is the number of zeros in the
eigenfunction. The ground state, $n=0$, has no zeros: the first excited state
is odd and has one zero at the origin and the second excited state is even
and has two zeros. Thus the oscillatory parts of the solution
are represented by orbits that
encircle the phase-space origin $(n+1)/4$ times before approaching the
origin almost parallel to the $y'$-axis. There are $n/4$ oscillations
in the interval $0\leq x \leq \xt$ so we have the approximate relation
$\xt=nT/4$. For later use it is convenient to introduce the
scaled variables
\[
N=\frac{\pi}{2}\omega n, \quad \sE=\frac{E}{N}, \quad {\rm and} \quad
z=\frac{2E}{Aa^{2}}\geq 2,
\]
in terms of which $k^{2}=1/(z-1)$ and the quantisation condition becomes
\drlabel{eq:2-9}
\begin{equation}
\omega \xt(\sE,z)=\sqrt{\frac{N}{2\sE}}g(z), \quad
g(z)=\frac{2K(k)}{\pi\sqrt{1-1/z}}.
\label{eq:2-9}
\end{equation}
For large $z$, $g(z)=1+\frac{3}{4z}+O(z^{-2})$.

Finally, we need an approximation to the motion for $x>\xt$. The value of
$y(\xt)$ must be close to the barrier height, $y(\xt)\simeq y_{m}(\xt)$: if
$y(\xt)\ll y_{m}(\xt)$ the orbit would complete another $\frac12$ period
and if $y(\xt)> y_{m}(\xt)$ it would escape. But if $y(\xt)\simeq y_{m}(\xt)$
the required subsequent orbit is approximated by expanding about the point
in phase space that follows the potential maximum, by making the canonical
transformation
\[
y=Q+y_{m}(x), \quad \frac{dy}{dx}=P+\frac{dy_{m}}{dx}
\]
and expanding the equations of motion to second-order. Then if $\xze>0$
is the time $\Eb(\xze)=0$ for $\xt <x <\xze$ the equations of motion are
\[
\frac{dQ}{dx}=P, \quad \frac{dP}{dx}= \Eb(x)Q-\frac{d^{2}y_{m}}{dx^{2}},
\quad \frac{d^{2}y_{m}}{dx^{2}}=-\frac{E\omega^{2}}{2\Eb(x)\sqrt{A\Eb(x)}}.
\]
These equations may be solved numerically and it is seen that $Q(x)$
remains small provided both $|P(\xt)|$ and $|\Eb(\xt)Q(\xt)-y_{m}''(\xt)|$
are small or zero. As $x\to\xze$ the solution diverges. However, over the
interval of interest this expansion shows that an approximate
solution is
\drlabel{eq:2-11}
\begin{equation}
y(x) \simeq y_{m}(x)=\sqrt{\frac{\Eb(x)}{A}}, \quad \xt \leq x < \xze,
\quad \Eb(\xze)=0.
\label{eq:2-11}
\end{equation}
This is, of course, the standard Thomas-Fermi approximation, obtained
from equation~\ref{eq:1-1} by ignoring the kinetic energy term.

Some idea of the accuracy of the approximations~\ref{eq:2-10} and~\ref{eq:2-11}
is given in the next figure comparing these with an exact solution. In this
case $E=15$, $A=100$ which gives $a=0.23976$ and $\xt=2.7272$.

\begin{center}
\begin{picture}(240,120)
 \put(0,0){\epsfig{file=hass02.eps}}
 \put(120,63){$x$}
 \put(170,50){$x_{0}$}
 \put(120,30){$\xt$}
 \put(28,105){$y(x)$}
 \put(118,38){\vector(-1,1){20}}  
\end{picture}

\refstepcounter{figure}
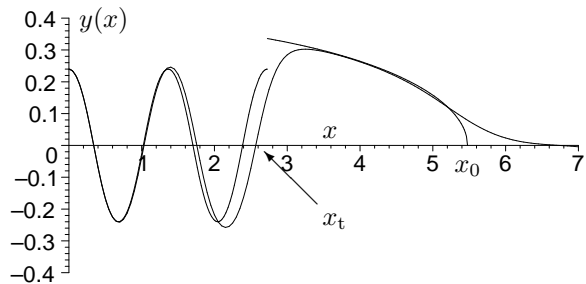
\drcaption{Graphs of the exact solution, the Thomas-Fermi
solution~\ref{eq:2-11} for $\xt\leq x\leq \xze$ and the adiabatic
solution~\ref{eq:2-10} for $0\leq x \leq \xt$ in the case
$E=15$ and $A=100$, for which $a=0.23976$ and $\xt=2.7272$.\label{f:02}}
\end{center}

\noindent
In the next section we use equations~\ref{eq:2-2}, \ref{eq:2-5}
and~\ref{eq:2-10} to approximate the eigenvalues of equation~\ref{eq:1-1}
and to obtain an approximate scaling law.

\section{An approximate scaling law}
\drlabel{sec:scal}\label{sec:scal}
Here we show that the approximations described above may be used to
derive an approximate scaling law relating the
energy, $E$, quantum number $n$ and the nonlinearity parameter $A$ by the
single equation,
\drlabel{eq:3-1}
\begin{equation}
\sE=H\left( \omega A N^{-3/2}\right),
\quad \sE=\frac{2E}{\pi n\omega}=\frac{E}{N}
\label{eq:3-1}
\end{equation}
for some function $H$.  A consequence of this is that the energy levels
behave like those of the linear oscillator in that the difference
$E_{n+1}(A)-E_{n}(A)$ is almost indepenent of $n$ and also of $A$.

In order to derive this relation we first express
$z$ in terms of $\sE$ using the adiabatic and the quantisation conditions,
equations~\ref{eq:2-6} and~\ref{eq:2-9} respectively. These equations may
be combined to give
\drlabel{eq:3-2}
\begin{equation}
\frac{2\sqrt{2}}{3\pi} \left( 1- \frac{g(z)^{2}}{4\sE^{2}} \right)^{3/2}=
\frac{F(k)}{z}, \quad k^{2}=\frac{1}{z-1}
\label{eq:3-2}
\end{equation}
which, in principle gives $z(\sE)$. The behaviour of this function is
shown in the next figure where $1/z$ is plotted as a function of $\sE=E/N$.

\begin{center}
\begin{picture}(240,120)
 \put(0,0){\epsfig{file=hass03.eps}}
 \put(145,20){$\sE$}
 \put(28,105){$1/z$}
\end{picture}

\refstepcounter{figure}
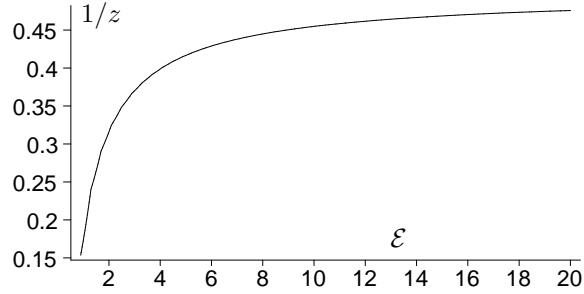
\drcaption{Graph of $1/z(\sE)$.\label{f:03}}
\end{center}

\noindent
As $z\to\infty$, $k\to 0$, $g\to 1$ and $F\to 1$, and so $2\sE\to 1$: in
this limit,
\[
\frac{1}{z}=\frac{2\sqrt{2}}{3\pi} \frac{(4\sE^{2}-1)^{3/2}}{(2\sE)^{3}},
\quad 2\sE \sim 1.
\]
As $\sE$ increases $1/z(\sE)$ increases monotonically to $1/2$.

The normalisation condition, equation~\ref{eq:2-2}, can be written in the form
\drlabel{eq:3-4}
\begin{equation}
1=2na^{2}\int_{0}^{T/4}dx\,{\rm sn}(\tau,k)^{2} +
2\int_{\xt}^{\xze}dx\,\frac{\Eb(x)}{A}.
\label{eq:3-4}
\end{equation}
The first of these integrals may be evaluated using relations given in
Abramowitz and Stegun (1965, section~16.25), so we have
\drlabel{eq:3-5}
\begin{equation}
1=\frac{2na^{2}}{\sqrt{2E-Aa^{2}}}\,\frac{K(k)-E(k)}{k^{2}} +
\frac{2}{3\omega A}(2E)^{3/2} -
\frac{\xt}{3A}\left( 6E-\omega^{2}\xt^{2} \right).
\label{eq:3-5}
\end{equation}
In terms of the scaled variables introduced in equation~\ref{eq:2-9} this
becomes
\drlabel{eq:3-6}
\begin{equation}
\frac{3A\omega}{2N^{3/2}}=(2\sE)^{3/2} \left\{ 1+
\frac{3}{4\sE} \left( \frac{4}{\pi}\frac{K(k)-E(k)}{k^{2}\sqrt{z(z-1)}}-
g(z)\right)+\frac{g(z)^{3}}{16\sE^{3}} \right\}.
\label{eq:3-6}
\end{equation}
Since $k^{2}=1/(z-1)$ and $z$ is a function of $\sE$ through
equation~\ref{eq:3-2}, the right hand side of this equation depends only
upon $\sE$. Thus $\sE$ is a function only of the variable $\omega AN^{-3/2}$,
which is the scaling law~\ref{eq:3-1}.

This analysis can be carried further with more approximations, but first
we show the graph of the ratio
\drlabel{eq:3-7}
\begin{equation}
R(\sE)=\frac{3A\omega}{2N^{3/2}} \,\frac{1}{(2\sE)^{3/2}}
\label{eq:3-7}
\end{equation}
which is seen from equation~\ref{eq:3-6}, and the fact that $z\to 2$,
tends to unity as $\sE\to\infty$.

\begin{center}
\begin{picture}(240,120)
 \put(0,0){\epsfig{file=hass04.eps}}
 \put(150,38){$\sE$}
\end{picture}

\refstepcounter{figure}
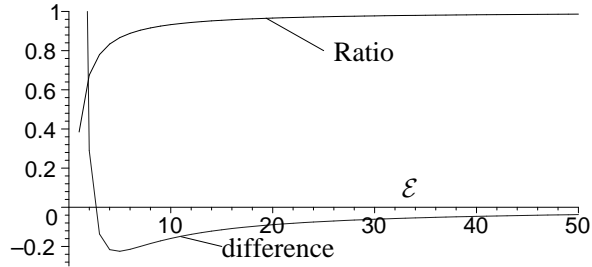
\drcaption{Graphs of the ratio $R(\sE)$,
equation~\ref{eq:3-7}, and the difference
$100(R(\sE)-R_{1}(\sE))$.\label{f:04}}
\end{center}

\noindent
Expanding equation~\ref{eq:3-6} in powers of $1/z$ gives
\drlabel{eq:3-8}
\begin{equation}
 \frac{3A\omega}{2N^{3/2}}=(2\sE)^{3/2}
\left\{1-\frac{3}{4\sE}\left( 1-\frac{1}{4z}+\cdots\right)+\frac{1}{16\sE^{3}}
\left(1+\frac{9}{4z}+\cdots \right) \right\}.
\label{eq:3-8}
\end{equation}
An analysis of $R(\sE)$ suggest that $1-R(\sE)\sim \sE^{-1}$ for large $\sE$,
that in this range $z\simeq \frac12$ and that $z$ changes relatively slowly
with $\sE$. Thus a simple approximation to this ratio is given by
setting $z$ equal to its asymptotic value, $z=2$, to give
\[
R(\sE)\simeq R_{1}(\sE)=1-\frac{21}{32\sE}.
\]
The graph of $100(R(\sE)-R_{1}(\sE))$ is shown in
figure~\ref{f:04} and this demonstrates the accuracy of this simple
approximation.

On using $R_{1}$ to approximate $R(\sE)$ in equation~\ref{eq:3-6} and
rearranging the equation we obtain
\drlabel{eq:3-9}
\begin{equation}
E_{n}(A)=\frac12 \left( \frac{3A\omega}{2}\right)^{2/3} +
\frac{7\pi}{32}\omega n +{\rm higher\;order\;terms}.
\label{eq:3-9}
\end{equation}
The first term in this equation is just the Thomas Fermi approximation, which
follows from the normalisation condition, equation~\ref{eq:3-4}, by
setting $\xt=0$. The second term increases linearly with $n$ and, because
the trap potential quadratic, is independent of $A$. Higher-order corrections
come from the expansion about the asymptotic value of $z$ and are complicated
and not warranted because of other approximations made.

The scaling law~\ref{eq:3-1} exists because the trap potential is homogeneous
in $x$, so the adiabatic condition~\ref{eq:2-6} may be expressed in terms
of only two variables. For the quadratic potential these are
$\sE=\frac{2E}{\pi\omega n}=\frac{E}{N}$ and $z=\frac{2E}{Aa^{2}}$ and it
is the form of these variables that gives the scaling law~\ref{eq:3-1} and
ultimately the energy level~\ref{eq:3-9}. If the trap potential is
$(\omega x)^{2p}/2p$, $p\geq 1$, the scaled energy may be taken to be
$\sE=2EN^{-2p/(2p+1)}$ and then the scaling law~\ref{eq:3-1} becomes
\[
E=N^{\frac{2p}{2p+1}}H\left( \frac{A\omega}{N^{\frac{2p+1}{p+1}}}\right)
\]
and the energy levels become
\drlabel{eq:3-10}
\begin{equation}
E_{n}(A)=\frac{1}{2p}
\left( \left( p+\frac12\right)A\omega \right)^{\frac{2p}{2p+1}} +
\frac{7\pi\omega n}{32\sqrt{p}}
\left( \left( p+\frac12\right)A\omega\right)^{\frac{p-1}{2p+1}}.
\label{eq:3-10}
\end{equation}
When $p=1$ this reduces to equation~\ref{eq:3-9}, but when $p\neq 1$ the
coefficient of $n$ depends upon the nonlinearity, $A$.

\section{Variational method}
\drlabel{sec:var}\label{sec:var}
Yukalov {\em et al} (1997) have used  re-normalised perturbation theory
to obtain analytic approximations to the energy levels of the $3d$ nonlinear
Schr\"{o}dinger equation. Here we show that this method is equivalent to
a Euler-Lagrange variational method and that the resulting energy levels
of the excited states do not satisfy the scaling law described in
equation~\ref{eq:3-9}. Thus this method cannot be as accurate as implied
by Yukalov {\em et al} (1997).

With the Lagrangian
\drlabel{eq:4-1}
\begin{equation}
L(y,y',x)=\frac12 \left( \frac{dy}{dx}\right)^{2} +
\frac12 \omega^{2}x^{2}y^{2} + \frac12 A y^{4}
\label{eq:4-1}
\end{equation}
and treating the energy as the Lagrange multiplier we see that the
Euler-Lagrange equations with the functional and the constraint
\[
\overline{J}[y]=\int_{-\infty}^{\infty} dx\,\left[ L(y,y',x)-Ey^{2} \right],
\quad \int_{-\infty}^{\infty}dx\,y(x)^{2}=1,
\]
gives equation~\ref{eq:1-1}, with $\mu=\hbar=1$, and that the energy is then
given by
\drlabel{eq:4-2}
\begin{equation}
E=\int_{-\infty}^{\infty} dx\,\left[ \frac12 \left( \frac{dz}{dx}\right)^{2} +
\frac12 \omega^{2}x^{2}z^{2} + A z^{4} \right]
\label{eq:4-2}
\end{equation}
where $z(x)$ is a solution of the Euler-Lagrange equation. For trial
functions satisfying the normalisation condition we may use the simpler
functional
\drlabel{eq:4-3}
\begin{equation}
J[y]=\int_{-\infty}^{\infty}dx\,L(y,y',x).
\label{eq:4-3}
\end{equation}

A natural trial function is
\drlabel{eq:4-4}
\begin{equation}
z(x)=\sqrt{\frac{a}{h_{n}}}\,H_{n}(ax)\exp\left( -\frac12 a^{2}x^{2}\right),
\quad h_{n}^{2}=2^{n}\,n!\,\sqrt{\pi}
\label{eq:4-4}
\end{equation}
where $a$ is the variational parameter. Then the functional~\ref{eq:4-3}
becomes
\drlabel{eq:4-5}
\begin{equation}
J(a)=\frac12\left( n+\frac12\right)\left( a^{2}+\frac{\omega^{2}}{a^{2}}\right)
+\frac{aA}{2h_{n}^{2}}I_{n},
\quad I_{n}=\int_{-\infty}^{\infty}dw\,H_{n}(w)^{4}e^{-2w^{2}}.
\label{eq:4-5}
\end{equation}
This is stationary so the appropriate value of $a$ is given by the positive
root of
\drlabel{eq:4-6}
\begin{equation}
\frac{\omega^{2}}{a^{3}}=a+\frac{AI_{n}}{(2n+1)h_{n}^{2}}
\quad {\rm and\;then}\quad
E_{n}=\frac12 \left( n+\frac12 \right)
\left( a^{2} + \frac{\omega^{2}}{a^{2}}\right) + \frac{aA}{h_{n}^{2}}I_{n}.
\label{eq:4-6}
\end{equation}

If $A=0$ these equations give the unperturbed energy levels
and if $A$ is small perturbation theory may be used to obtain the equivalent
of Yukalov {\em et al} (1997), equation~44. For $A\gg 1$ and $n=0$ they give
$E_{0}=0.677 (\omega A)^{2/3}$ which is 3.4\% larger than the Thomas-Fermi
energy, given by the first term in equation~\ref{eq:3-9}. In this limit
of large $A$ perturbation theory may be used to give
\drlabel{eq:4-8}
\begin{equation}
E_{n}=\frac54 (2n+1)\omega^{2}B^{2/3} \left( 1+\frac{\epsilon}{15}+
\frac{\epsilon^{2}}{15}+\cdots\right),
\quad B=\frac{AI_{n}}{(2n+1)\omega^{2} h_{n}^{2}}, \quad
\epsilon=\frac{1}{\omega^{2}B^{4/3}}.
\label{eq:4-8}
\end{equation}

It is also clear from equations~\ref{eq:4-6} that $E/N$ depends only
upon the variable $z=AI_{n}/((2n+1)h_{n}^{2}\sqrt{\omega})$, which is
different from the scaling law derived in the previous section.

\section{Numerical results}
\drlabel{sec:num}\label{sec:num}
In this section we compare the behaviour of the energy levels of
equation~\ref{eq:1-1}, computed numerically, with the predictions of the
above formula, equations~\ref{eq:3-9} and~\ref{eq:4-6}.

One method of numerically solving equation~\ref{eq:1-1} is to perform
a two-dimensional search in the $(a,E)$ plane, where $E$ is the energy and
for even solutions $y(0)=a>0$ and for odd solutions $y'(0)=a>0$. These
solutions must {\em a}) satisfy the quantisation condition, {\em b})
tend to zero as $x\to\infty$ and {\em c}) satisfy the normalisation condition.
Since most solutions are unbounded this
calculation is expedited by using a good first approximation, which is
given by
\[
\tilde{y}(x)=\left\{ \begin{array}{cc}
(a+xy_{m}(\xt)/\xt)\cos\Omega x, & 0 \leq x \leq \xt \\
y_{m}(x), & \xt \leq x \leq \xze \\
0, & x > \xze=\sqrt{2E}
\end{array} \right.
\]
where $y_{m}(x)$ is the Thomas-Fermi solution defined in
equation~\ref{eq:2-11} and $\Omega=2\pi/T$ where $T$ is the period defined in
equation~\ref{eq:2-7}. In practice the harmonic balance approximation
$\Omega^{2}=2E-2a^{2}A/2$ was used for $\Omega$. The oscillatory part of
this approximation has a slowly increasing amplitude in order that
$\tilde{y}(x)$ is continuous at $x=\xt$.

This approximation has two free parameters, $a$ and $E$, which were varied
using the Marquardt algorithm to find values that simultaneously satisfied the
normalisation condition~\ref{eq:2-2} and the quantisation
condition~\ref{eq:2-9}. For $A=200$ this crude approximation gives
a relative error of less than 1\% for the ground state and 5\% for the
$16^{{\rm th}}$ energy level.

In the second stage of the calculation we use the energy $E$ found above
and vary $a$ to
find a value at which $|y(\xf)|<\delta$, for some small $\delta$
and where $\xf=1.25\xze$. This was achieved using a shooting algorithm
that that varied $a$ according to the value of $y(\xf)$. The solution
obtained in this manner is not normalised, but we find that for small
changes in $E$, $\int_{0}^{\xf}dx\,y(x)^{2}$ depends approximately
linearly on $E$ so it is possible to interpolate the energy to obtain
values of $(a,E)$ that give a correctly normalised solutions.

In the following table are shown energy levels for $A=100$ and $200$. The
exact numerical values are well approximated by the straight lines
$E_{n}\simeq 14.04 +0.66n$ and $E_{n}\simeq 22.40+0.74n$, for $A=100$
and $200$ respectively,
and the gradient of these lines is close to that predicted by
equation~\ref{eq:3-9}. The energy levels of the variational method do
not behave in this manner, particularly for large $A$, and we conclude
that the excited energy levels
given by the re-normalised perturbation method used by Yukalov {\em et al}
(1997) is not accurate for the one-dimensional nonlinear
Schr\"{o}dinger equation.

\begin{center}
\begin{tabular}{|c|cccc|} \hline
\multicolumn{5}{c}{$A=100$} \\ \hline
$n$ & 0 & 2 & 4 & 6 \\ \hline
$E_{n}$ (numerical) & 14.02 & 15.37 & 16.69 & 17.98 \\
$E_{n}$ (equation~\ref{eq:3-9}) & 14.12 & 15.47 & 16.84 & 18.20 \\
$E_{n}$ (equation~\ref{eq:4-6}) & 14.60 & 18.70 & 20.34 & 21.69 \\
\hline\hline
\multicolumn{5}{c}{$A=200$} \\ \hline
$n$ & 0 & 2 & 4 & 6 \\ \hline
$E_{n}$ (numerical) & 22.42 & 23.87 & 25.34 & 26.86 \\
$E_{n}$ (equation~\ref{eq:3-9}) & 22.41 & 23.77 & 25.13 & 26.49 \\
$E_{n}$ (equation~\ref{eq:4-6}) & 23.17 & 29.53 & 31.77 & 33.34 \\
\hline\hline
\end{tabular}
\end{center}

\section{Conclusions}
We have shown that the energy levels $E_{n}$ of the Gross-Pitaevskii
equation~\ref{eq:1-1} satisfy the approximate scaling law~\ref{eq:3-1}, which
relates the variables $E,\,n,\,\omega,\,A$ in a single equation, which leads
to the approximate energy levels~\ref{eq:1-2}. We have shown that other
homogeneous trap potentials lead to similar scaling laws but only the
energy levels of the quadratic trap have a coefficient of $n$ that is
independent of the nonlinear constant, see equation~\ref{eq:3-10}. It is also
shown that the energy levels of the re-normalised perturbation method
of Yukalov {\em et al} (1997) are equivalent to a simple variational
method and do not satisfy the scaling law derived here.

The method used to derive these results involves interpreting the
Gross-Pitaevskii equation as a mechanical system with a slowly varying
potential, so that the idea of adiabatic invariance can be used. With this
equivalence the spatial coordinate becomes the time, so the generalisation
to the $2d$- or $3d$ Gross-Pitaevskii equation is not apparent. For symmetric,
many dimensional systems, however a similar approach may be possible though
there are some problems with singularities at the origin that need to be
resolved.

\subsubsection*{Acknowledgements}
We thank Drs J A Vaccaro and O Steuernagel for helpful discussions.

\subsubsection*{References}
\begin{trivlist}
\item[] Abramowitz  M and Stegun I A 1965 {\em Handbook of Mathematical
functions} (Dover)
\item[] Friedrich H 1998 {\em Theoretical Atomic Physics} Springer
\item[] Percival I C  and Richards D 1982 {\em Introduction to Dynamics}
(Cambridge University Press)
\item[] Yukalov V I, Yukalova E P and Bagnato V S 1997 Phys Rev {\bf A56}
4845--54
\end{trivlist}
\end{document}